# A MACHINE LEARNING-BASED APPROACH TO CATEGORIZE RESEARCH JOURNALS

Rabia Shabbir Ranjha [a], Arshad Ali, PhD [b,*], Shahid Yousaf [a]

[a] *Department of Computer Science, University of Lahore, 1-km Defence road near Bhuptian Chowk, Lahore, Pakistan*

[b] *FAST School of Computing, NUCES, Lahore Block B Faisal Town, Lahore, 54770, Pakistan*

In this modern technological era, categorization and ranking of research journals is gaining popularity among researchers and scientists. It plays a significant role for publication of their research findings in a quality journal. Although, many research works exist on journal categorization and ranking, however, there is a lack of research works to categorize and predict the journals using suitable machine learning (ML) techniques.

This work aims to categorize and predict various academic research journals. This work suggests a hybrid predictive model comprising of five steps. The first step is to prepare the dataset with twenty features. The second step is to pre-process the dataset. The third step is to apply an appropriate clustering algorithm for categorization. The fourth step is to apply appropriate feature selection techniques to get an effective subset of features. The fifth step involves some ensemble plus non-ensemble methods to train the model. The model is trained on a full set of features, and a selected subset of features is obtained by applying three feature selection techniques. After model training, the prediction results are evaluated in terms of precision, recall, and accuracy. The results can help the researchers and the practitioners in predicting the journal's category.

## 1. Introduction

In this modern and advanced academic age, the researchers are producing more and more research work in different domains of academia such as medical, business, natural phenomenon, social sciences, and physical sciences. Researchers of various domains face the same kind of challenge while publishing their research work in a suitable journal managed by authorized publishers. Mainly there are two methods for journal ranking in a suitable manner. Subjective method, also known as manual, perceptual or peer-reviewed method, enables the experts or committee of any institute to create list of categorized journals. While, objective method, also known as bibliometric method. These metrics (bibliometric indicators) are quantitative measures and help to assess the research output. These metrics play an important role in finding research trends in different areas and categorizing journals using some common indicators such as count, citation, h-index, and impact factor etc.

Ho Initially, the journals categorization was done by a manual method providing institutions list prepared by scholarly experts or committee votes. This method of categorization was criticized at those times because it is biased by the individual background involved in preparing the journals categorization list [1]. The results by this method did not show the reasonable status and the actual quality of the journals. This method also creates the problem of differences between the humans involved in the journal's categorization within the institute.

Hence, this measure proved to be an inaccurate method. Consequently, journal categorization done by the experts, institutes, and communities have turned into the bibliometric indicator method, which is independent of humans [2].

Apart from above discussion, the bibliometric indicators are of two types: 1) sole indicator, The importance of bibliometric indicators is varying from journal to journal according to their fields. Impact factor (IF) is a commonly used features to categorize the journal, but a single feature is not enough to decide. In comparison, the multiple features make it possible to ensure the standing of a journal in a specific academic domain. 2) multiple indicators, Different indicators give different categorizations according to several features of the journals. Most of the time, a set of indicators provides different categorizations. Meanwhile, another set of indicators gives other categorizations for the same journals. The single indicator categorization is complex because of the complexity of various journals. The single indicator may not provide adequate results, whereas the multiple indicators offer more adequate results.

*Corresponding author: arshad.ali1@nu.edu.pk

Research on journal categorization is still in its initial phases. The research gap in finding the journal's category can be overcome by applying various Machine Learning (ML) techniques. ML-based techniques help in categorizing the journals and to train a reliable predictive model in an efficient way to predict the journal's category. For this, first step involves preparation of dataset that contains the record of journals. Second step is about pre-processing of data. Preprocessing involves data cleaning, data transformation. Third, the clustering technique to categorize the data correctly. Fourth, we used three feature selection techniques on the clustered dataset. Fifth step is application of classification techniques on whole dataset and on featured dataset achieved by applying feature selection. All these are possible through ML-based techniques.

## 2. Literature Review

The academic journal is a platform for researchers to publish their findings [3]. These academic journals may be reviewed by manual method, which is also called peer review method. In this method journals are reviewed by the panel of experts. This method of categorization was criticized at those times because it is biased by the individual background involved in preparing the journals categorization list [1] The results by this method did not show the reasonable status and the actual quality of the journals. Hence, this measure proved to be an inaccurate method.

Besides manual method, the automated method is more reliable which is free from the human decisions. These bibliometric features involve IF, h-index, journal homepage, full title, five-years impact factor, eigen-factor score, publisher, SNIP, website, cite-score, citations, acceptance rates, no. of issues per year, coverage, total articles, SJR, country, cited half-life, immediacy-factor, article influence score, age, open access, ISSN, and percentile. Impact factor is a commonly used features to categorize the journal, but a single feature is not enough to decide. In comparison, the multiple features make it possible to ensure the standing of a journal in a specific academic domain. So, there is a demand for a reliable evaluation method based on automation that is independent of humans and takes multiple features in its consideration to specify a journal in its specified domain. We also going to apply this method in research study.

We need to apply different ML techniques to categorize the journals in a good manner. On prepared dataset, Preprocessing is the most essential and first step to perform machine learning experiments. Preprocessing involves data cleaning, data transformation, and data reduction. In this regard, we applied data cleaning to remove the unnecessary feature, data transformation to transform the data on a single scale between 0 and 1.

After preprocessing, unsupervised learning plays an essential role in several fields. In this regard, researchers use different clustering techniques. For example, k-means and k-medoids with different distance formulas. Wu et al. in 2018 [4] applied k-means along with Euclidean distance formula as an unsupervised learning to cluster the data. Bektas et al. in 2019 [5] uses k-medoids along with Gower distance formula to cluster the data.

Beside clustering, the validation of clusters is most essential to validate either the clustering performed accurately or not. In this regard, several research studies utilized Silhouette width and Adjusted Rand Index (ARI). Lengyel et al. in 2019 [6] used Silhouette width to validate cluster results on Iris dataset (a flowers dataset). Somasundaram et al. in 2011 [7] used ARI to validate the clusters.

After clustering validation, Feature selection techniques often claims to produce better prediction results. It is a practice in which the features of dataset are analyzed to get features having strong correlation with the class variable, rest features get reduced from the dataset. Chandrashekar et al. in 2014 [8] proposed a survey on the importance of feature selection methods. Xu et al. in 2017 [9] uses CFS-Subset evaluator to get the effective to predict cardiovascular heart patients. Juneja et al. in 2020 [10] had also used chi-square to rank the feature of breast cancer patients. We also applied these three methods to get important features.

Classification is the next step to train a predictive model that helps to classify the journals. Model can be trained with Percentage Split Method (PSM) and k-fold Cross Validation Method (CVM). Several literatures utilized the k-fold

CVM to predict the several diseases [11, 12, 13]. Hussain et al [14] have presented a research study to predict diabetes patients using both PSM and 10-fold CVM. We also performed experiments using both. In this innovative age, researchers are using ML-based classification techniques in several domains. Ishaq et al [15] used nine classifiers DT, AdaBoost, LR, SGD, RF, GBM, NB, ETC, and SVM. Halim et al [16] have proposed a predictive model on journal categorization and prediction by applying some preprocessing, clustering, feature selection techniques, and classification techniques (K-Nearest Neighbors).

Rest of the paper is organized as complete overview of the proposed approach or methodology applied in this research study to categorize the journals, involves the results of each step of proposed methodology, includes the conclusion of the proposed study consisting of several experiments, conducted by fixing multiple challenges in building a reliable predictive model. It also concludes that which machine learning techniques remain fruitful in predicting the accuracy.

### 3. Proposed Methodology

This section talks about the design of the proposed methodology which is divided into three phases. Figure 1 shows the complete overview of the proposed methodology. The proposed model consists of following steps:

- Preprocessing
- Clustering and Feature Selection
- Classification Techniques and Model Evaluation

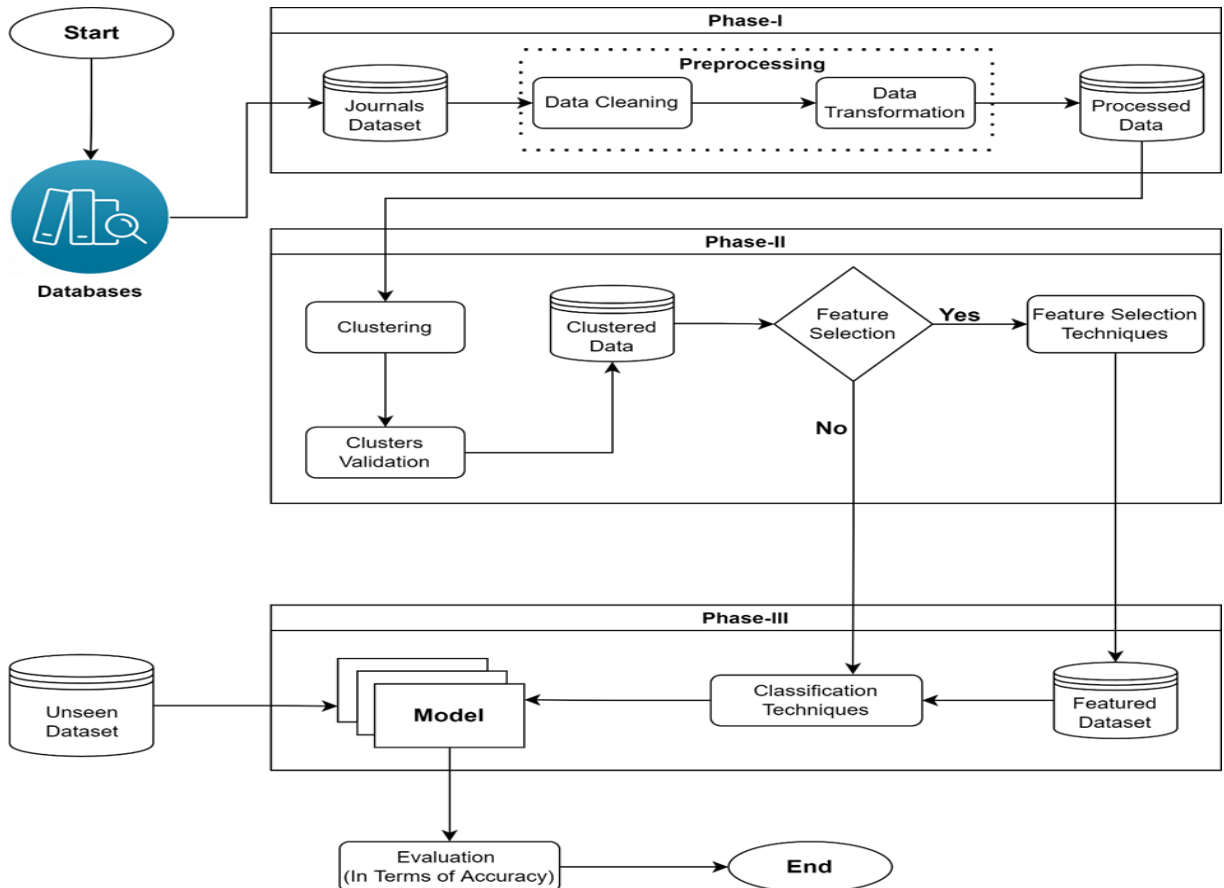

*Figure. 1: Proposed Methodology*

### *3.1. Data Description*

This section shows the data description, takes the input of journals data from multiple data stores such as Scimago, Research Gate, Scopus, Academic Accelerator, Editage, Guide2Research, MedSci, and Science Direct. The dataset consists of twenty features, shown Table 1. After preparing the dataset, the next task is to perform preprocessing to shape the data into suitable form.

Table 1: The Feature to be Used in Suggested Model

| Features | Description |
| --- | --- |
| Full Title | It includes the full name of the research journal. |
| ISSN | It includes the international standard serial number of the journal. |
| Publisher | A platform that prepares and issue books, journals, and music for sales. |
| Journal Impact Factor | Average number of citations of published paper in a particular journal |
| Journal homepage | A place where you find the online availability of journals and related information. |
| Website | It represents the journals website address. |
| Cite Score | CS is measured by yearly average citation for journal |
| SJR | SJR (Scimago Journal Ranking) is the average no. of weighted citations received in a year that is published in a journal in the last three years. |
| SNIP | SNIP (Source Normalized Impact per Paper), It is calculated as the no. of citations given in the present year to publication in the past three years is divided by the total no. of publication in the last three years. |
| Country | Define the origin of journal. |
| Coverage | Coverage represents the starting year of a journal, like when the journal emerged. |
| Hirsch Index | The h-index captures research output based on the total number of publications and the total number of citations to those works. |
| Eigen Factor Score | The Eigen-factor Score measures the number of times articles from the journal published in the past five years have been cited in the Journal Citation Reports (JCR) year. |
| Article Influence Score | Measures the average influence, per article, of the papers published in a journal. Calculated by dividing the Eigen-factor by the number of articles published in the journal. |
| Immediacy Score/index | The Immediacy Index is the average number of times an article is cited in the year it is published. |
| Cited-half life | The Cited Half-Life is the median age of a journal's articles that were cited in the JCR year |
| Total articles | Total articles publish in journal during year |
| Open access | Access of journal is open of published article |
| 5 years impact factor | The 5-year Impact Factor is the average number of times articles from the journal published in the past five years have been cited in the JCR year. |
| No. of issues per year | No of issues pointed in an article during a specific year. |

### 3.2. Pre-processing

Preprocessing involves data cleaning and data transformation. Data cleaning is applied to remove the unnecessary features. Data transformation is to transform categorical feature values into numeric values. Int this regard, there are two methods; one is the manual method and other is the automated method. We utilized both methods on our dataset. By applying these methods, the categorical features also become continuous. After that, we also applied scaling on transformed data to scale the values of each feature on a same scale between 0 and 1. We applied min max normalization to transform the values of each feature on a same scale. After applying preprocessing techniques, we obtained a preprocessed data which is then passed as an input to the phase-II.

### 3.3. Clustering

There are several techniques to cluster the instances of the dataset. As each feature of the preprocessed data is normalized on a same scale. So, k-medoid with Gower distance formula is best suited to cluster such kind of dataset. So, we utilized k-medoids to cluster the data. After performing clustering technique, the clustered dataset is passed for clusters validation. For this, we utilized Silhouette Width and Adjusted Rand Index (ARI) techniques to validate the clusters. After cluster validation the clustered dataset is passed for feature selection.

### 3.4. Feature selection

It is a machine learning technique that used to select best features from datasets and makes datasets lightweight. Feature selection is a process in which a subset of features from a wide variety of features is extracted that helps to get more reliable prediction results. We Applied three methods for feature selection:

**CFS-Subset-Eval Method:** CFS-Subset-Eval plus Best-First as a searching strategy is considered more important to achieve a suitable subset of features. So, we used the CFS-Subset-Eval as a filter and best-first as a searching strategy in this proposed study.

**Chi2 Method:** It is one of the most effective methods under filter methods and claims to produce better results for feature selection. So, we applied this method to get the effective number of features. In Chi2, we select the desire no of features and it gives the most highly correlated features. Firstly, we gave input five and get five most important features from 15 features then produces seven, ten, and twelve features as well.

**Random Forest Classifier:** We also applied individual feature ranking using RF classifier under embedded method to get the importance of each feature with the class variable. It gives the importance of each feature in terms of frequency values. Firstly, we select top five features with highest feature importance for classification. We also taken top seven, ten, and twelve features to check the accuracy on different sets of features.

After applying feature selection techniques, a new significant featured data is obtained as an output from phase-II and passed as an input to the phase-III. If feature selection is not applied, then the control directly goes to phase-III.

### 3.5. Classification techniques

It includes the classification techniques that help to form a reliable predictive model. We used nine classifiers using PSM and CVM with selected and all features. We use some non-ensemble and some ensemble-based classification techniques to train the model. Non-ensemble methods have a single classifier to make decisions, whereas ensemble methods include more than one classifier to make the decision. In this work, NB and MLP under non-ensemble method and RF, Ada-boost, Bagging, CB, ETC, L-GBM, and XGB under ensemble methods are chosen to classify.

### 3.6. Evaluation measures

After constructing predictive models from nine classifiers, an evaluation measure is necessary to perform on the prediction results. In this study, we perform three evaluation measures such as precision (PR), recall (RE), and accuracy (ACC). The mathematical representation of PR, RE, and ACC can be defined as follows:

$$\text{Precision (PR)} = \frac{(TP)}{(TP + FP)}$$

$$\text{Recall (RE)} = \frac{(TP)}{(TP + FN)}$$

$$\text{Accuracy (ACC)} = \frac{TP + TN}{TP + TN + FP + FN}$$

## 4. Results and Discussion

This chapter involves the statistical results of each experiment of the proposed research. This involves the results of each step of proposed methodology, also includes a discussion section on the results of several experiments with the experiments performed in this study and with the previous research work.

### 4.1. Results

This section shows the results of experiments for each phase. Several experiments are conducted in python using various libraries in order to classify the data to produce better prediction results.

#### 4.1.1. Preprocessing Results

In preprocessing of data, we applied data cleaning that removed five features from the dataset and those are as "Full Title, Website, Homepage, ISSN, and Total Articles". Full title and ISSN are representing the identity of a journal and identity does not play any role in prediction because prediction requires the statistical, numeric, or categorical values. Same scenario is in case of website and homepage that represents the URL links, again not important in prediction. Rest is the Total articles; it is important for prediction, but we removed this feature because the data against this feature in not available for most of the journals.

In terms of data normalization, we applied manual method on two features; one is the "Open Access" and other is the coverage. If a journal has open access, we considered it as 1 and if a journal has no open access, considered as 0. In case of coverage area, we manually count the number of years that a particular journal giving coverage from how many years. We also applied label encoding as an automation-based method on two features to transform the multiple categories into numeric; one is the Publisher and other is the Country. After that we applied data scaling to transform the values of all features on a same scale between 0 and 1.

After applying data cleaning and data transformation, we achieved a preprocessed data which is then passed as an input for clustering and feature selection.

### 4.1.2. Clustering Results

We applied k-medoids as a clustering algorithm along with Gower distance. We passed three as the value of K because we need to categorize our instances into three classes. First class represents the best quality journal, second for average, and third for the least. Now, two questions raised here; One is the clusters are valid? Second is about the category of clusters like which cluster (0, 1, and 2) belongs to which category, either best, average, or least? We used two cluster validation method to answer the first question. Silhouette validation result values between 0 and 1 shows that the clusters are valid, and it is more valid as the value is as close to 0. In our case, Silhouette showed the validation results as 0.238 which lies between the validation range and much closer to 0. Hence our clustering is valid. On the other hand, we used ARI. The ARI value 1 shows that clusters are valid and in our case its value is 1. Hence, second method also proved that our clusters are valid. Now to answer the second question, we consider the average values of each feature under each cluster and performed an analysis to answer the second question. Halim et al. also performed the same scenario to evaluate the suitable category of journals [16]. As shown in Figure 2,*Figure* cluster_1 represents the best category journals, cluster_0 represent the average, and cluster_2 represents the least.

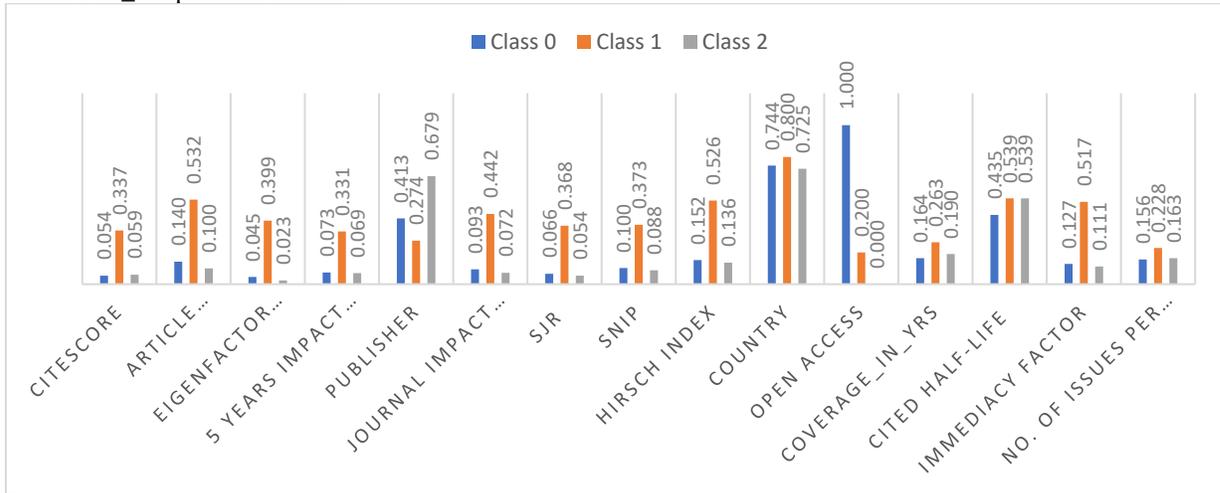

*Figure 2: The Analysis of K-Medoids Clustering on each feature*

### 4.1.3. Feature Selection Results

We applied three (CFS-Subset-Eval, RF, and Chi2) feature selection methods. RF returns the importance of each feature individually. Whereas Chi2 and CFS returns just the required number of features. The feature importance achieved by RF is shown in Table 2. CFS-Subset-Eval produces three features only; Open Access, Publisher, and Cited-Half Life. Later the features obtained by each techniques used separately for classification.

### 4.1.4. Classification Results

We perform classification using both PSM and CVM with nine classifiers (NB, MLP, Bagging, RF, XGB, CB, L-GBM, ETC, and AdaBoost). The experiments have been done on all features and selected features either by chi-square, RF, or CFS-Subset-Eval. First, we have discussed the prediction results taken with selected features than we have discussed the prediction results with a complete set of features. Let us have a brief look on each experiment with selected or with a complete set of features.

Table 2: Feature Importance of each Feature by RF

| Sr. No. | Feature Name | Feature Importance |
|---|---|---|
| 1 | No. of issues per year | 0.01356 |
| 2 | Country | 0.01476 |
| 3 | Article Influence Score | 0.01766 |
| 4 | Coverage in years | 0.01809 |
| 5 | Cited Half-Life | 0.02233 |
| 6 | Eigenfactor Score | 0.02315 |
| 7 | Immediacy factor | 0.02545 |
| 8 | SNIP | 0.03205 |
| 9 | 5 years impact factor | 0.03407 |
| 10 | Hirsch index | 0.0408 |
| 11 | CiteScore | 0.0442 |
| 12 | SJR | 0.04513 |
| 13 | Journal Impact Factor | 0.06027 |
| 14 | Publisher | 0.11633 |
| 15 | Open access | 0.4921 |

I. *Evaluation results with three features selected by CFS-Subset-Eval using 10-Fold CVM*

The model is trained and tested with nine classifiers on three features selected features by CFS-Subset-Eval using CVM as a training method. The subset of three features includes Open Access, Publisher, and Cited Half Life. The accuracy results obtained by each classifier are shown in Table 3. In comparison of accuracy with CVM, ETC outperform others with an accuracy value of 0.911 along with 0.606 PR, and 0.667 RE. Ada-Boost shows the least accuracy of 0.88 along with 0.685 PR, and 0.699 RE. The performance comparison of accuracy results against each classifier is shown in Figure 3.

Table 3: Evaluation Results with Three Features Selected by CFS-Subset-Eval using 10-Fold CVM

| Classifiers | PR | RE | ACC |
|---|---|---|---|
| NB | 0.606 | 0.667 | 0.911 |
| MLP | 0.606 | 0.667 | 0.911 |
| Bagging | 0.74 | 0.736 | 0.884 |
| RF | 0.753 | 0.765 | 0.894 |
| XGB | 0.718 | 0.732 | 0.888 |
| CB | 0.696 | 0.718 | 0.887 |
| L-GBM | 0.747 | 0.741 | 0.893 |
| ETC | 0.606 | 0.667 | 0.911 |
| AdaBoost | 0.685 | 0.699 | 0.88 |

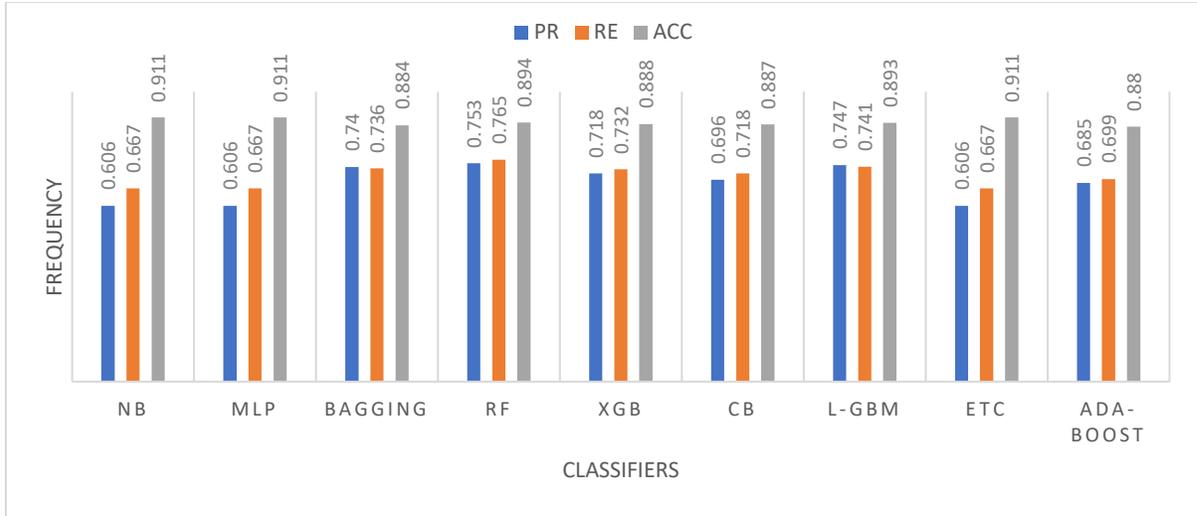

*Figure 3: Evaluation Results with Three Features Selected by CFS-Subset-Eval using 10-Fold CVM*

*II.    Evaluation results with three features selected by CFS-Subset-Eval using PSM*

The model is trained and tested with nine classifiers on three features selected by CFS-Subset-Eval using PSM as a training method. The accuracy results obtained by each classifier are shown in Table 4. In comparison of accuracy with PSM, NB outperform others with an accuracy value of 0.941 along with 0.627 PR, and 0.777 RE. Ada-Boost shows the least accuracy of 0.881 along with 0.732 PR, and 0.937 RE. The performance comparison of accuracy results against each classifier is shown in Figure 4.

Table 4: Evaluation Results with Three Features Selected by CFS-Subset-Eval using PSM

| Classifiers | PR | RE | ACC |
| --- | --- | --- | --- |
| NB | 0.627 | 0.777 | 0.941 |
| MLP | 0.627 | 0.777 | 0.941 |
| Bagging | 0.733 | 0.733 | 0.882 |
| RF | 0.755 | 0.748 | 0.897 |
| XGB | 0.732 | 0.739 | 0.882 |
| CB | 0.786 | 0.673 | 0.868 |
| L-GBM | 0.686 | 0.673 | 0.878 |
| ETC | 0.604 | 0.777 | 0.912 |
| AdaBoost | 0.732 | 0.739 | 0.881 |

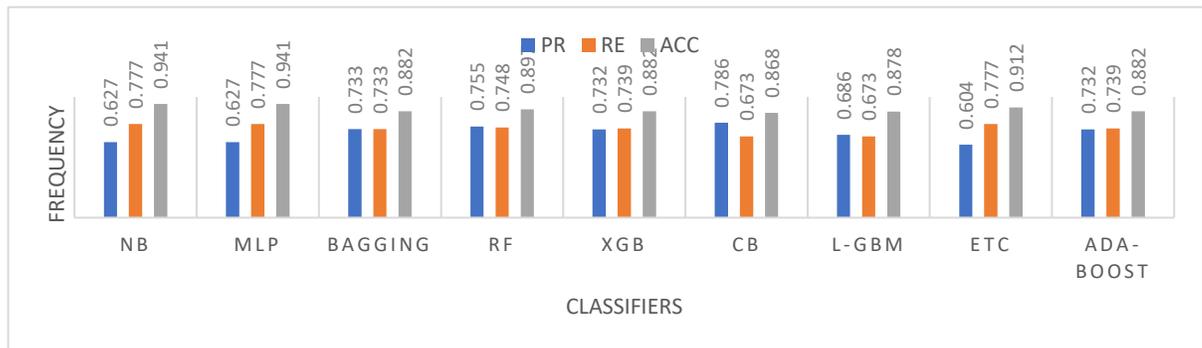

*Figure 4: Evaluation Results with Three Features Selected by CFS-Subset-Eval using PSM*

III. *Evaluation results with Chi2 ranked features using 10-Fold CVM*

The model is trained and tested with nine classifiers with Chi2 ranked features using 10-fold CVM as a training method. The accuracy, precision, and recall results obtained by each classifier are shown in Table 5, Table 6 and Table 7 respectively.

Table 5: Accuracy Values with Chi2 Ranked Features using 10-Fold CVM

| Classifiers | 5 Features | 7 Features | 10 Features | 12 Features | 15 Features |
| --- | --- | --- | --- | --- | --- |
| NB | 0.957 | 0.959 | 0.963 | 0.963 | 0.96 |
| MLP | 0.966 | 0.976 | 0.979 | 0.984 | 0.982 |
| Bagging | 0.964 | 0.969 | 0.96 | 0.969 | 0.961 |
| RF | 0.978 | 0.978 | 0.978 | 0.979 | 0.981 |
| XGB | 0.975 | 0.969 | 0.963 | 0.964 | 0.963 |
| CB | 0.975 | 0.978 | 0.966 | 0.976 | 0.97 |
| L-GBM | 0.969 | 0.969 | 0.966 | 0.975 | 0.973 |
| ETC | 0.985 | 0.982 | 0.982 | 0.987 | 0.987 |
| AdaBoost | 0.561 | 0.563 | 0.547 | 0.585 | 0.587 |

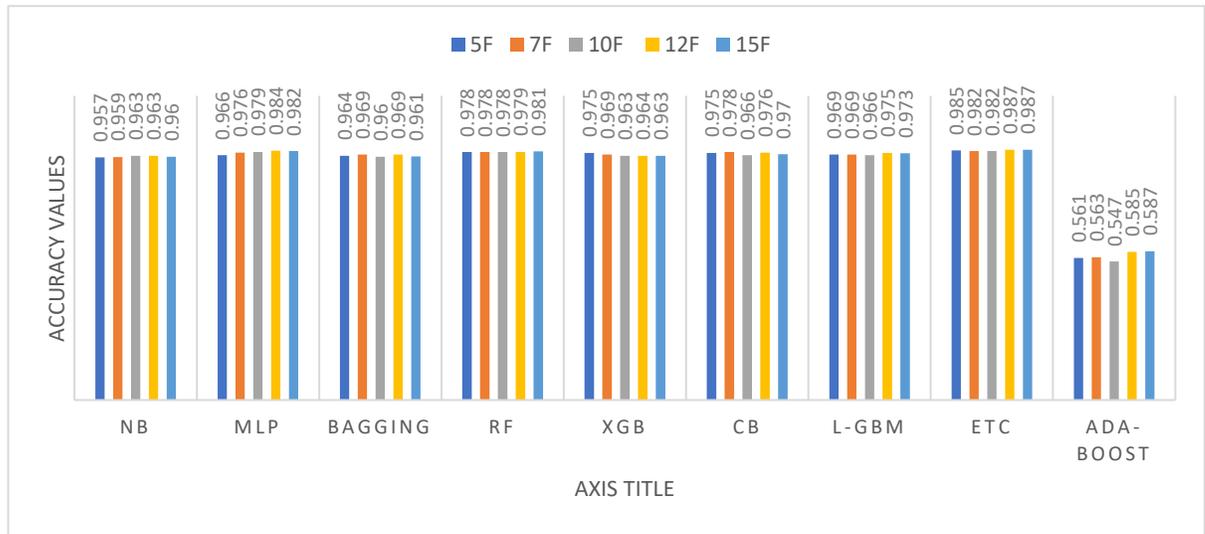

*Figure 5: Accuracy Comparison with Chi2 Ranked Features using 10-Fold CVM*

In comparison of evaluation measures, ETC outperform others with a maximum accuracy of 0.987 along with 0.98 and 0.992 precision values and the recall values as 0.95 and 0.948 by 12 and 15 features. In case of 7 and 10 features, it produces minimum of 0.982 accuracy along with 0.969, 0.98 precision and 0.932, 0.948 recall values, respectively. Whereas AdaBoost produces the least accuracy 0.547 along with 0.686 precision and 0.645 recall by 10 features, and maximum it produces 0.587 accuracy along with 0.698 precision and 0.663 recall by all features. The comparison of accuracy, precision, and recall values is shown in Figure 5, Figure 6 and Figure 7 respectively.

Table 6: Precision Values with Chi2 Ranked Features using 10-Fold CVM

| Classifiers | 5 Features | 7 Features | 10 Features | 12 Features | 15 Features |
|---|---|---|---|---|---|
| NB | 0.914 | 0.908 | 0.924 | 0.924 | 0.919 |
| MLP | 0.895 | 0.94 | 0.944 | 0.965 | 0.978 |
| Bagging | 0.933 | 0.932 | 0.924 | 0.929 | 0.944 |
| RF | 0.946 | 0.974 | 0.96 | 0.967 | 0.967 |
| XGB | 0.948 | 0.945 | 0.917 | 0.924 | 0.921 |
| CB | 0.953 | 0.963 | 0.914 | 0.974 | 0.958 |
| L-GBM | 0.938 | 0.944 | 0.94 | 0.961 | 0.96 |
| ETC | 0.979 | 0.969 | 0.98 | 0.98 | 0.992 |
| AdaBoost | 0.721 | 0.694 | 0.686 | 0.698 | 0.698 |

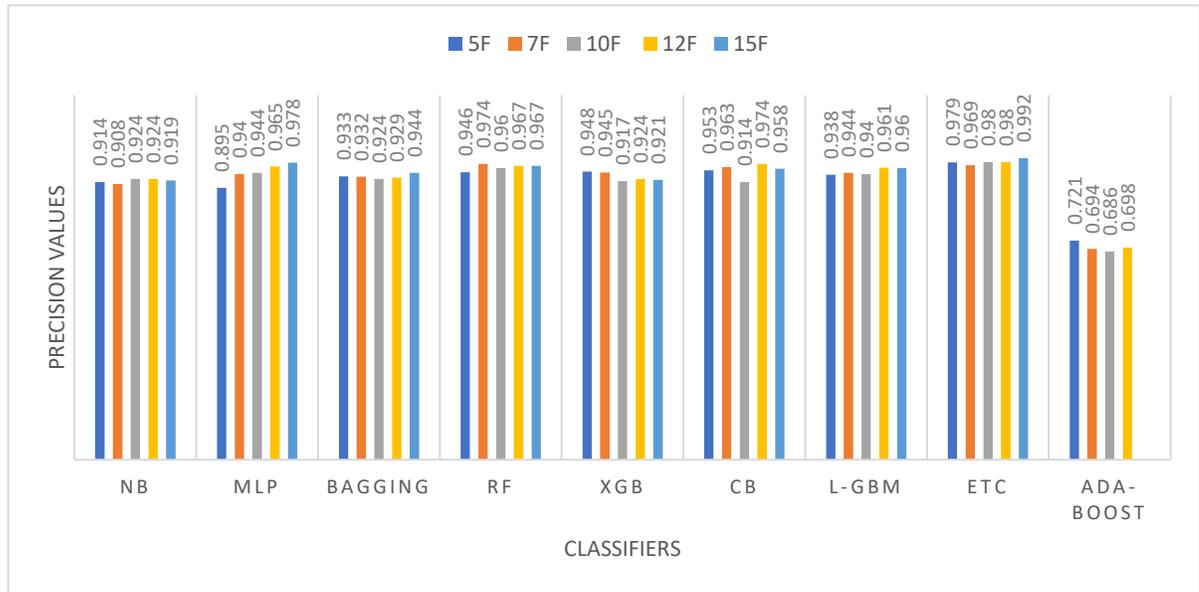

*Figure 6: Precision Comparison with Chi2 Ranked Features using 10-Fold CVM*

Table 7: Recall Values with Chi2 Ranked Features using 10-Fold CVM

| Classifiers | 5 Features | 7 Features | 10 Features | 12 Features | 15 Features |
|---|---|---|---|---|---|
| NB | 0.915 | 0.933 | 0.959 | 0.959 | 0.957 |
| MLP | 0.867 | 0.933 | 0.922 | 0.939 | 0.928 |
| Bagging | 0.91 | 0.896 | 0.91 | 0.902 | 0.901 |
| RF | 0.92 | 0.935 | 0.931 | 0.935 | 0.933 |
| XGB | 0.933 | 0.915 | 0.902 | 0.894 | 0.893 |
| CB | 0.918 | 0.929 | 0.9 | 0.929 | 0.907 |
| L-GBM | 0.929 | 0.937 | 0.913 | 0.928 | 0.922 |
| ETC | 0.938 | 0.932 | 0.948 | 0.95 | 0.948 |
| AdaBoost | 0.659 | 0.659 | 0.645 | 0.662 | 0.663 |

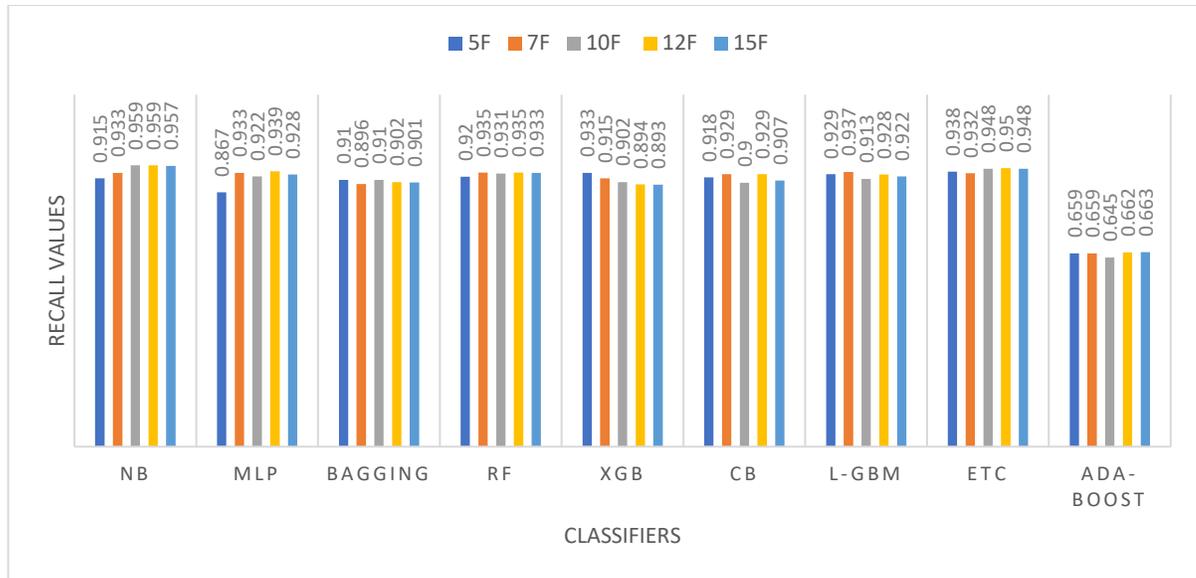

*Figure 7: Recall Comparison with Chi2 Ranked Features using 10-Fold CVM*

IV.     *Evaluation results with Chi2 ranked features using PSM*

The model is trained and tested with nine classifiers with Chi2 ranked features using PSM as a training method. The accuracy, precision, and recall results obtained by each classifier are shown in Table 8, Table 9 and Table 10 respectively.

In comparison of evaluation measures, ETC outperform others with a maximum accuracy of 0.985 along with precision values as 0.991, 0.991, 0.991, and 0.98 and the recall values as 0.944, 0.833, 0.952, and 0.958 by 10, 12, and 15 features, respectively. In case of 7 features, it produces minimum of 0.912 accuracy along with 0.93 precision and 0.918 recall values. Whereas AdaBoost produces the least accuracy 0.471 along with 0.794 precision and 0.612 recall by 10 features, and maximum it produces 0.633 along with 0.843 precision and 0.74 recall by all features. The comparison of accuracy, precision, and recall is shown in Figure 8, Figure 9 and Figure 10 respectively.

Table 8: Accuracy Values with Chi2 Ranked Features using PSM

| Classifiers | 5 Features | 7 Features | 10 Features | 12 Features | 15 Features |
| --- | --- | --- | --- | --- | --- |
| NB | 0.941 | 0.985 | 0.912 | 0.912 | 0.966 |
| MLP | 0.966 | 0.912 | 0.985 | 0.985 | 0.966 |
| Bagging | 0.966 | 0.926 | 0.966 | 0.926 | 0.985 |
| RF | 0.971 | 0.912 | 0.966 | 0.971 | 0.971 |
| XGB | 0.971 | 0.985 | 0.985 | 0.966 | 0.971 |
| CB | 0.971 | 0.912 | 0.971 | 0.985 | 0.971 |
| L-GBM | 0.966 | 0.912 | 0.941 | 0.971 | 0.985 |
| ETC | 0.971 | 0.912 | 0.985 | 0.985 | 0.985 |
| AdaBoost | 0.588 | 0.588 | 0.471 | 0.617 | 0.633 |

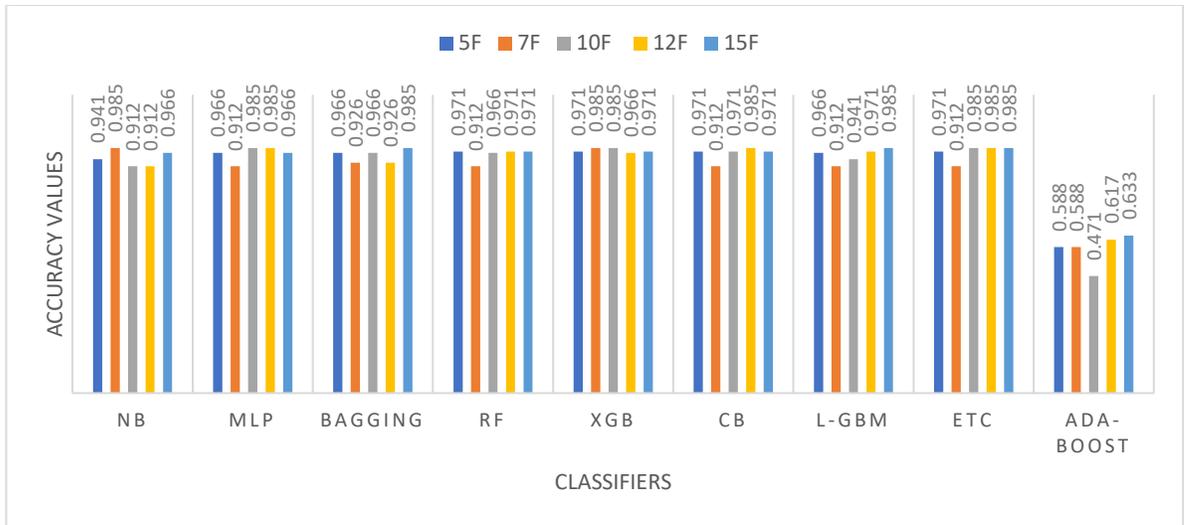

Figure 8: Accuracy Comaprison with Chi2 Ranked Features using PSM

Table 9: Precision Values with Chi2 Ranked Features using PSM

| Classifiers | 5 Features | 7 Features | 10 Features | 12 Features | 15 Features |
| --- | --- | --- | --- | --- | --- |
| NB | 0.918 | 0.952 | 0.857 | 0.846 | 0.91 |
| MLP | 0.966 | 0.937 | 0.986 | 0.99 | 0.923 |
| Bagging | 0.946 | 0.947 | 0.909 | 0.853 | 0.989 |
| RF | 0.977 | 0.937 | 0.909 | 0.943 | 0.958 |
| XGB | 0.977 | 0.987 | 0.963 | 0.923 | 0.958 |
| CB | 0.988 | 0.937 | 0.933 | 0.99 | 0.958 |
| L-GBM | 0.946 | 0.937 | 0.899 | 0.942 | 0.963 |
| ETC | 0.988 | 0.937 | 0.963 | 0.958 | 0.98 |
| AdaBoost | 0.727 | 0.771 | 0.632 | 0.668 | 0.843 |

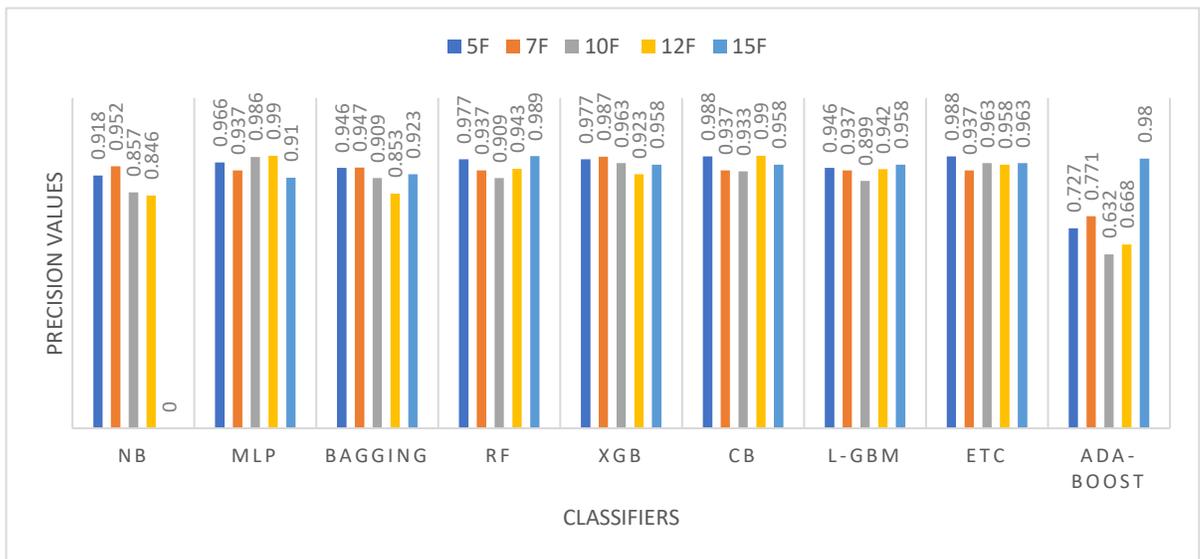

Figure 9: Precision Comparison with Chi2 Ranked Features using PSM

Table 10: Recall Values with Chi2 Ranked Features using PSM

| Classifiers | 5 Features | 7 Features | 10 Features | 12 Features | 15 Features |
| --- | --- | --- | --- | --- | --- |
| NB | 0.876 | 0.986 | 0.932 | 0.934 | 0.979 |
| MLP | 0.888 | 0.788 | 0.958 | 0.952 | 0.894 |
| Bagging | 0.914 | 0.815 | 0.986 | 0.873 | 0.985 |
| RF | 0.936 | 0.788 | 0.986 | 0.941 | 0.958 |
| XGB | 0.936 | 0.944 | 0.99 | 0.894 | 0.958 |
| CB | 0.936 | 0.788 | 0.984 | 0.952 | 0.958 |
| L-GBM | 0.924 | 0.788 | 0.958 | 0.942 | 0.989 |
| ETC | 0.936 | 0.788 | 0.99 | 0.988 | 0.958 |
| AdaBoost | 0.715 | 0.701 | 0.688 | 0.626 | 0.74 |

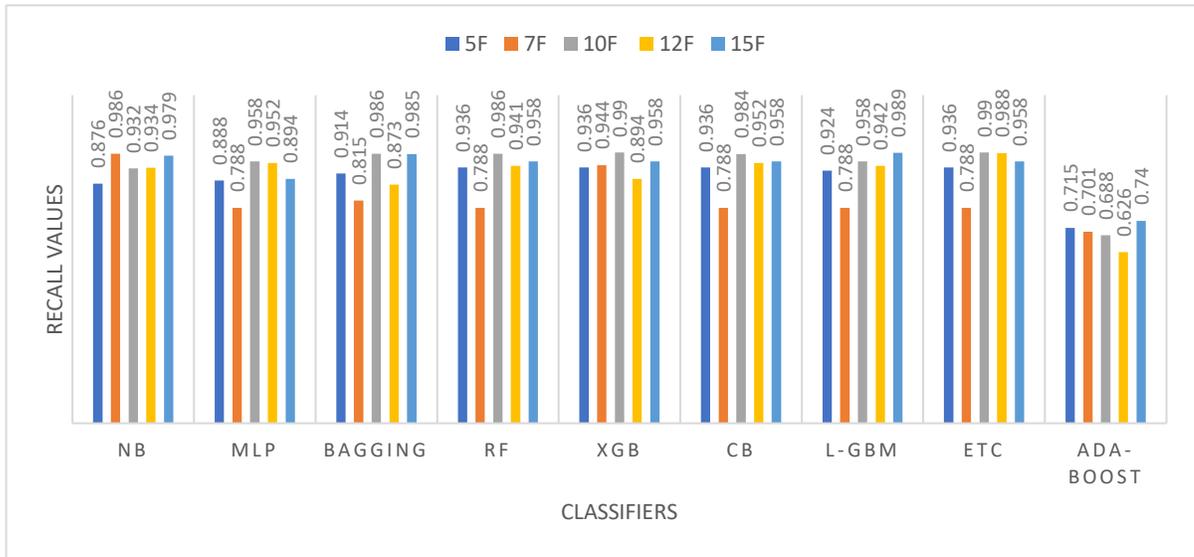

Figure 10: Recall Comparison with Chi2 Ranked Features using PSM

V. Evaluation results with RF ranked features using 10-Fold CVM

The model is trained and tested with nine classifiers with RF ranked features using 10-fold CVM as a training method. The accuracy, precision, and recall results obtained by each classifier are shown in Table 11, Table 12 and Table 13 respectively.

In comparison of evaluation measures, ETC outperform others with a maximum accuracy value of 0.987 along with precision values as 0.991 and 0.98 and the recall values as 0.952 and 0.958 by 12 and all features, respectively. In case of 5 features, it produces minimum of 0.975 accuracy, 0.93 precision and 0.918 recall. Whereas AdaBoost produces the least accuracy 0.552 along with 0.68, 0.977 precision and 0.692, 0.833 recall by 5 and 7 features, and maximum it produces 0.6 along with 0.843 precision and 0.774 recall by 12 features. The comparison of accuracy, precision, and recall are shown in Figure 11, Figure 12 and Figure 13 respectively.

Table 11: Accuracy Values with RF Ranked Features using 10-Fold CVM

| Classifiers | 5 Features | 7 Features | 10 Features | 12 Features | 15 Features |
|---|---|---|---|---|---|
| NB | 0.966 | 0.981 | 0.963 | 0.964 | 0.96 |
| MLP | 0.969 | 0.978 | 0.985 | 0.982 | 0.982 |
| Bagging | 0.963 | 0.96 | 0.96 | 0.958 | 0.961 |
| RF | 0.975 | 0.981 | 0.982 | 0.984 | 0.981 |
| XGB | 0.973 | 0.97 | 0.963 | 0.964 | 0.963 |
| CB | 0.972 | 0.976 | 0.972 | 0.976 | 0.97 |
| L-GBM | 0.973 | 0.978 | 0.972 | 0.972 | 0.973 |
| ETC | 0.975 | 0.982 | 0.984 | 0.987 | 0.987 |
| AdaBoost | 0.552 | 0.552 | 0.587 | 0.6 | 0.587 |

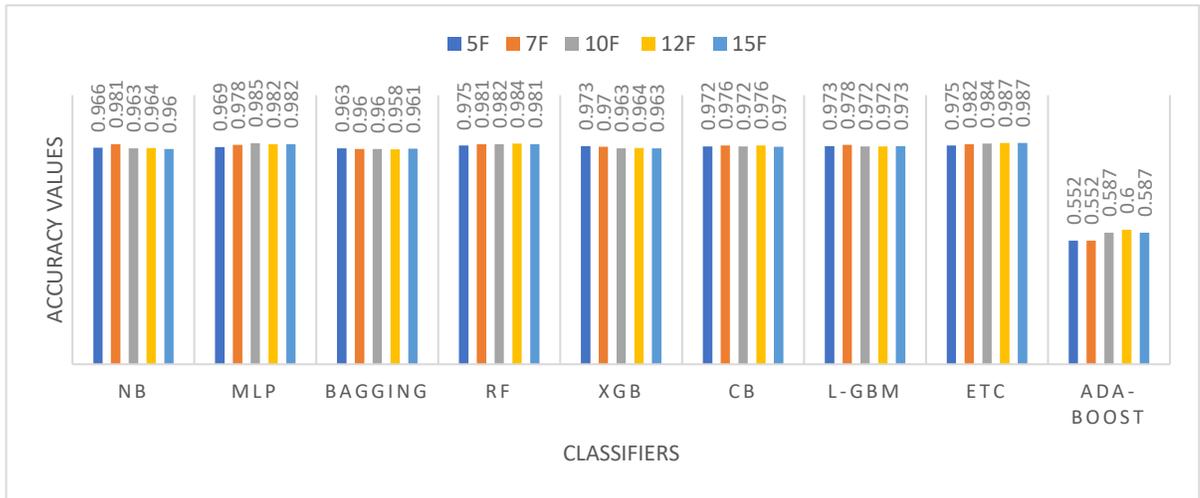

Figure 11: Accuracy Comparison with RF Ranked Features using 10-Fold CVM

Table 12: Precision Values with RF Ranked Features using 10-Fold CVM

| Classifiers | 5 Features | 7 Features | 10 Features | 12 Features | 15 Features |
|---|---|---|---|---|---|
| NB | 0.921 | 0.959 | 0.927 | 0.93 | 0.919 |
| MLP | 0.924 | 0.95 | 0.99 | 0.953 | 0.978 |
| Bagging | 0.91 | 0.91 | 0.941 | 0.924 | 0.944 |
| RF | 0.939 | 0.955 | 0.955 | 0.978 | 0.967 |
| XGB | 0.936 | 0.959 | 0.918 | 0.924 | 0.921 |
| CB | 0.933 | 0.964 | 0.938 | 0.966 | 0.958 |
| L-GBM | 0.943 | 0.96 | 0.927 | 0.956 | 0.96 |
| ETC | 0.94 | 0.971 | 0.971 | 0.988 | 0.992 |
| AdaBoost | 0.672 | 0.684 | 0.699 | 0.704 | 0.698 |

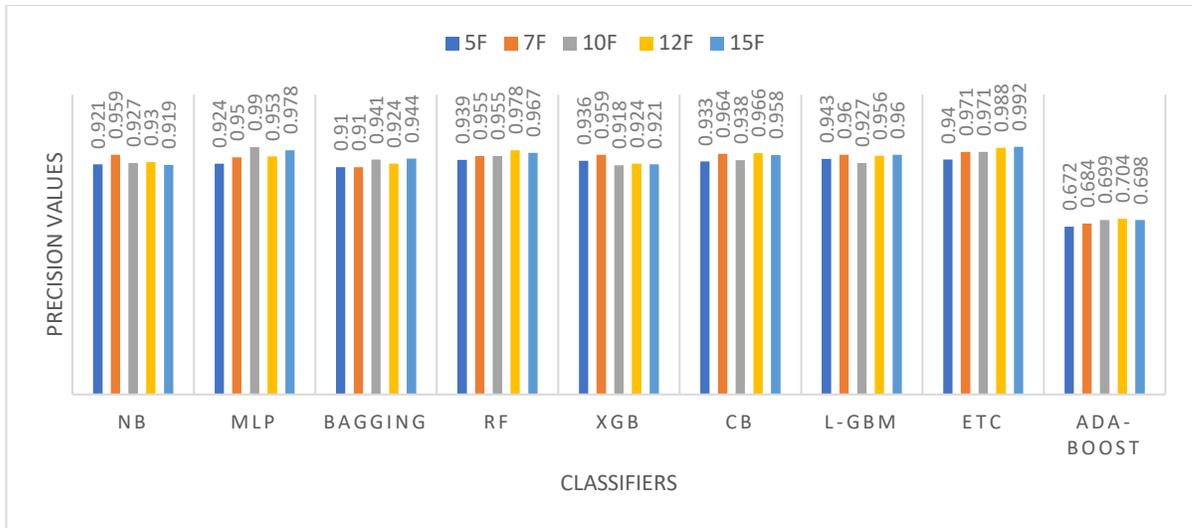

*Figure 12: Precision Comparison with RF Ranked Features using 10-Fold CVM*

Table 13: Recall Values with RF Ranked Features using 10-Fold CVM

| Classifiers | 5 Features | 7 Features | 10 Features | 12 Features | 15 Features |
| --- | --- | --- | --- | --- | --- |
| NB | 0.909 | 0.955 | 0.947 | 0.952 | 0.957 |
| MLP | 0.878 | 0.911 | 0.944 | 0.939 | 0.928 |
| Bagging | 0.912 | 0.91 | 0.899 | 0.92 | 0.901 |
| RF | 0.929 | 0.925 | 0.942 | 0.942 | 0.933 |
| XGB | 0.931 | 0.926 | 0.898 | 0.894 | 0.893 |
| CB | 0.93 | 0.925 | 0.908 | 0.925 | 0.907 |
| L-GBM | 0.932 | 0.94 | 0.947 | 0.921 | 0.922 |
| ETC | 0.931 | 0.952 | 0.948 | 0.95 | 0.948 |
| AdaBoost | 0.669 | 0.656 | 0.663 | 0.671 | 0.663 |

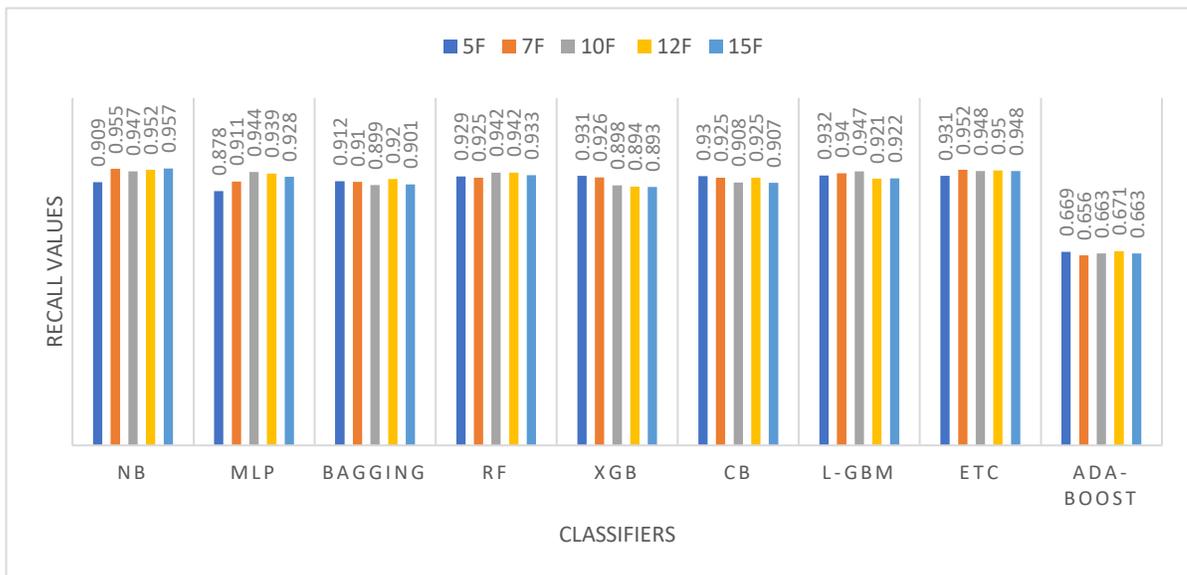

*Figure 13: Recall Comparison with RF Ranked Features using 10-Fold CVM*

## VI. Evaluation results with RF ranked features using PSM

The model is trained and tested with nine classifiers with RF ranked features using PSM as a training method. The accuracy, precision, and recall results obtained by each classifier are shown in Table 14, Table 15 and Table 16 respectively.

Table 14: Accuracy Values with RF Ranked Features using PSM

| Classifiers | 5 Features | 7 Features | 10 Features | 12 Features | 15 Features |
| --- | --- | --- | --- | --- | --- |
| NB | 0.941 | 0.985 | 0.947 | 0.985 | 0.966 |
| MLP | 0.966 | 0.985 | 0.985 | 0.985 | 0.966 |
| Bagging | 0.941 | 0.971 | 0.971 | 0.971 | 0.985 |
| RF | 0.971 | 0.971 | 0.966 | 0.985 | 0.971 |
| XGB | 0.971 | 0.985 | 0.971 | 0.971 | 0.971 |
| CB | 0.981 | 0.971 | 0.941 | 0.971 | 0.971 |
| L-GBM | 0.986 | 0.985 | 0.941 | 0.971 | 0.985 |
| ETC | 0.966 | 0.985 | 0.985 | 0.985 | 0.985 |
| AdaBoost | 0.496 | 0.618 | 0.602 | 0.632 | 0.633 |

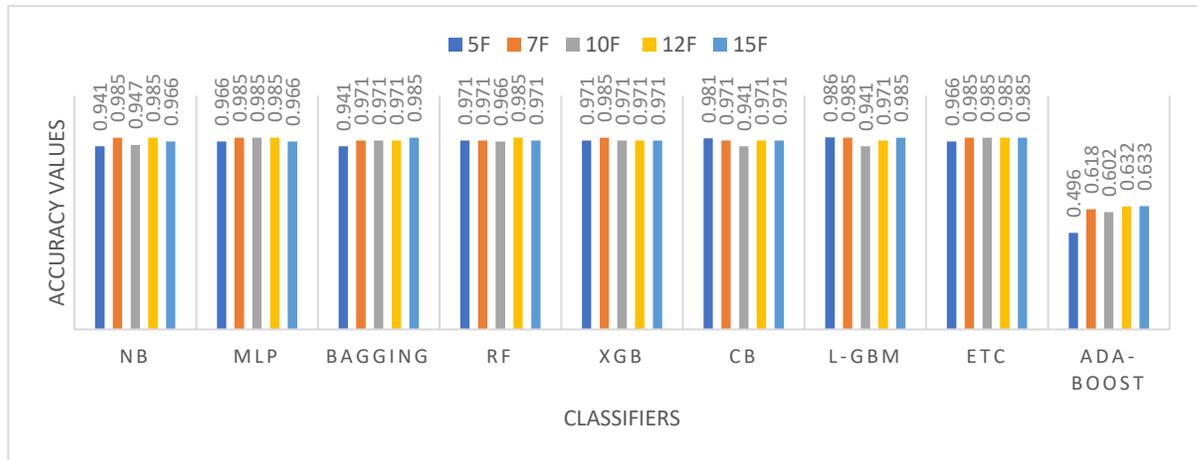

Figure 14: Accuracy Comparison with RF Ranked Features using PSM

In comparison of evaluation measures, ETC outperform others with a maximum accuracy value of 0.985 along with precision values as 0.963, 0.958, and 0.98 and the recall values as 0.99, 0.988, and 0.958 by each group of features except 5 features. MLP also produces the highest accuracy but slightly less precision and recall values. In case of 5 features, it produces minimum of 0.966 accuracy along with 0.937 precision and 0.788 recall. AdaBoost produces the least accuracy 0.496 along with 0.68 precision and 0.692 recall by 5 features, and maximum it produces 0.633 accuracy along with 0.843 precision and 0.74 recall by all features. The comparison of accuracy, precision, and recall is shown in Figure 14, Figure 15 and Figure 16 respectively.

Table 15: Precision Values with RF Ranked Features using PSM

| Classifiers | 5 Features | 7 Features | 10 Features | 12 Features | 15 Features |
| --- | --- | --- | --- | --- | --- |
| NB | 0.918 | 0.988 | 0.899 | 0.966 | 0.91 |
| MLP | 0.964 | 0.992 | 0.99 | 0.99 | 0.923 |
| Bagging | 0.893 | 0.98 | 0.825 | 0.947 | 0.989 |
| RF | 0.987 | 0.977 | 0.964 | 0.991 | 0.958 |
| XGB | 0.987 | 0.991 | 0.825 | 0.976 | 0.958 |
| CB | 0.99 | 0.977 | 0.953 | 0.947 | 0.958 |
| L-GBM | 0.991 | 0.991 | 0.953 | 0.947 | 0.963 |
| ETC | 0.93 | 0.991 | 0.991 | 0.991 | 0.98 |
| AdaBoost | 0.68 | 0.688 | 0.794 | 0.742 | 0.843 |

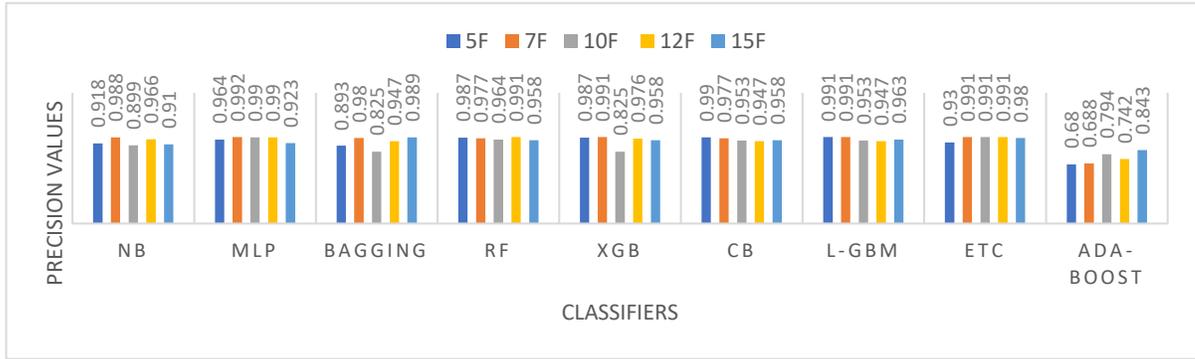

Figure 15: Precision Comparison with RF Ranked Features using PSM

Table 16: Recall Values with RF Ranked Features using PSM

| Classifiers | 5 Features | 7 Features | 10 Features | 12 Features | 15 Features |
| --- | --- | --- | --- | --- | --- |
| NB | 0.866 | 0.927 | 0.99 | 0.984 | 0.979 |
| MLP | 0.878 | 0.927 | 0.833 | 0.963 | 0.894 |
| Bagging | 0.893 | 0.833 | 0.825 | 0.947 | 0.985 |
| RF | 0.927 | 0.833 | 0.899 | 0.952 | 0.958 |
| XGB | 0.927 | 0.927 | 0.825 | 0.936 | 0.958 |
| CB | 0.969 | 0.833 | 0.852 | 0.947 | 0.958 |
| L-GBM | 0.969 | 0.927 | 0.852 | 0.947 | 0.989 |
| ETC | 0.918 | 0.944 | 0.833 | 0.952 | 0.958 |
| AdaBoost | 0.692 | 0.662 | 0.612 | 0.774 | 0.74 |

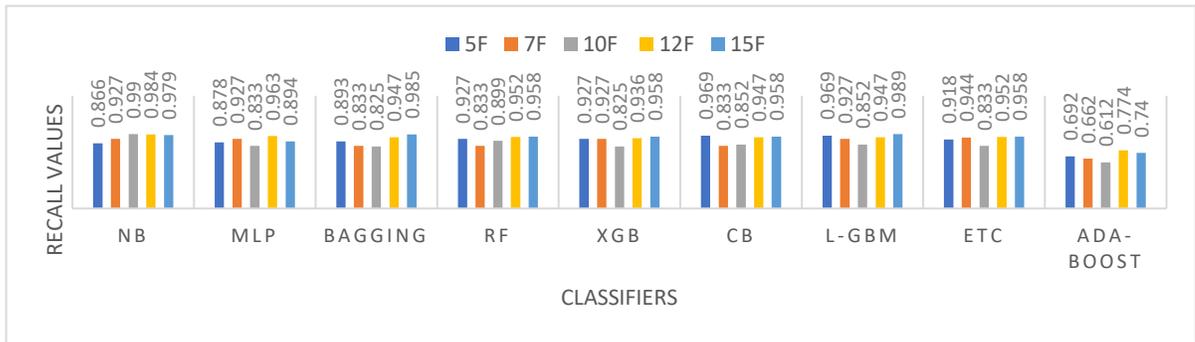

Figure 16: Recall Comparison with RF Ranked Features using PSM

### 4.1.5. Comparison of proposed approach with existing works

Table 17 provides comparison of our suggested approach with existing research works i.e., Halim et al. [16] and Feng et al. [17] in terms of evaluation measures.

Table 17: The performance comparison with Halim et al. [16] and Feng et al. [17]

| Literature | Preprocessing | Clustering | Classifier | Feature Selection | Model Training | ACC | PR | RE |
|---|---|---|---|---|---|---|---|---|
| Feng et al. [17] | No | - | KNN | 6 | - | 0.713 | - | - |
| Feng et al. [17] | No | - | BPNN | 6 | - | 0.47 | 0.45 | 0.48 |
| Halim et al. [16] Dataset | Yes | K-Means | KNN | 19 | PSM | 0.925 | 0.903 | 0.973 |
| Halim et al. [16] techniques on our Dataset | Yes | K-Means | KNN | MI | PSM | 0.985 | 0.983 | 0.952 |
| Proposed research | Yes | K-Medoids with Gower distance formula | ETC | Chi2 with 12 features | CVM | 0.987 | 0.98 | 0.95 |
| | | | | RF with 12 features | CVM | 0.987 | 0.988 | 0.95 |
| | | | | All features | CVM | 0.987 | 0.99 | 0.94 |

It is observed that the proposed approach outperforms with ETC along with CVM with highest accuracy value of 98.7% as shown in Table 17. Halim et al. uses K-means for clustering, KKN for classification, and Mutual Information for features selection and proposes 92.6% accuracy. We also try K-means and KNN on our dataset and it produces reliable accuracy value of 98.5%. So, our proposed approach is best for journals categorization and prediction. The comparison of evaluation measures is shown in Figure 17.

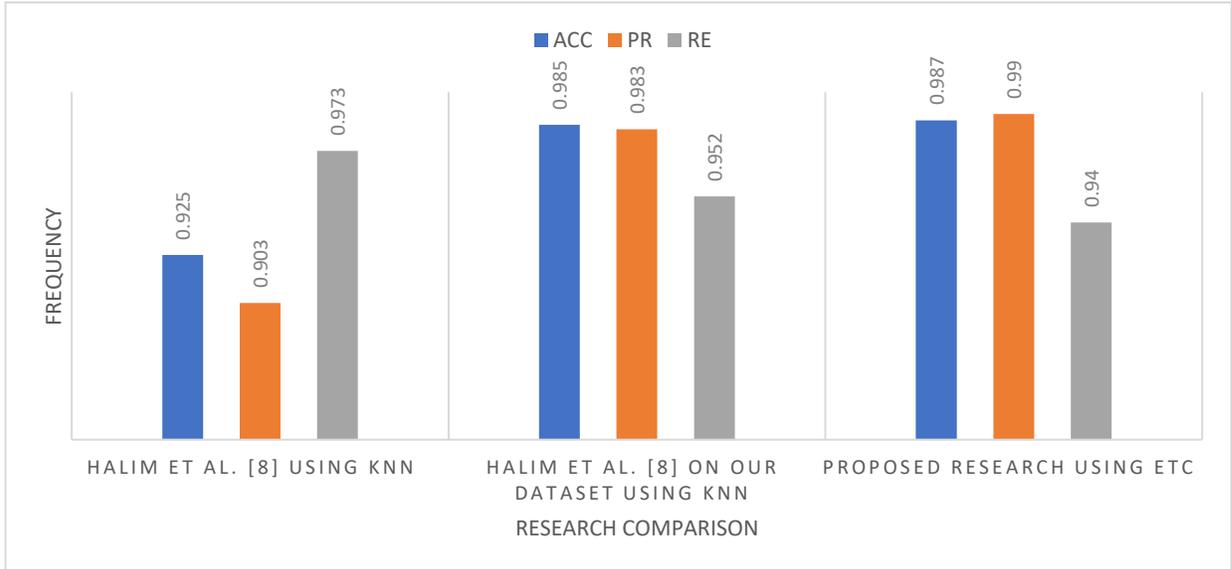

*Figure 17: The performance comparison with Halim et al.* [16].

*4.2. Discussion*

This section covers the discussion on several experiments that have been conducted in this study. The discussion consist of mainly three parts as follows:

Discussion on Feature Selection Techniques: Feature selection techniques often aim to produce fruitful results to find out suitable features that help in producing reliable prediction results. In this proposed study, we performed three feature selection techniques; one is the CFS-Subset-Eval and second is related the feature ranking by RF, third is Chi2. Chi2 and RF produces much reliable prediction results.

Discussion on Ensemble vs. Non-Ensemble Classifiers: We applied both ensemble and non-ensemble classifiers along with several experiments. The results showed that the ensemble classifiers outperform ever. In comparison of all ensemble classifiers, the ETC outperform others.

Discussion on Accuracy Results: There is a slight change in accuracy results achieved from the several experiments conducted in this study. We used nine classifiers on selected and all features with CVM and PSM as training methods. We achieved highest accuracy of 98.7% by using Extra-Tree-classifier with 12 features obtained by Chi2 and RF along with CVM.

**VII.    Conclusion and Future Work**

The proposed study conducted several experiments using suitable ML-based techniques to build a predictive model that can help to predict the category of research journals. In this regard, we prepared a dataset of journals, performed preprocessing, clustering, feature selection, and classification techniques. We achieved highest accuracy using all features with 10-fold CVM and ETC under ensemble classifiers mostly outperform with accuracy value of 98.7%. The results showed that ETC with 10-fold-CVM and 12 selected features by Chi2 with 98.7% outperform others with a smaller number of features.

There are two limitations in the proposed work. First, the dataset taken in this study is small. We planned to enlarge the size of dataset. Second is to develop a web or an android application that helps researchers to predict the category of research journal.